\documentclass[twocolumn,trackchanges]{aastex701}

\usepackage{xcolor}%

\begin{document}

\title{Diffusive or Ballistic? Distributions and Spectra of PeV Cosmic Rays around Microquasars}

\author[orcid=0000-0003-0058-9719,sname='Fujita']{Yutaka Fujita}
\affiliation{Department of Physics, Graduate School of Science, Tokyo Metropolitan University, 1-1 Minami-Osawa, Hachioji-shi, Tokyo 192-0397, Japan}
\email[show]{y-fujita@tmu.ac.jp}  

\author[orcid=0000-0002-3110-5732,sname='Takahashi']{Rohta Takahashi}
\affiliation{National Institute of Technology, Tomakomai College, Tomakomai 059-1275, Japan}
\email{takahashi@tomakomai-ct.ac.jp}

\author[orcid=0000-0001-8181-7511,sname='Kawanaka']{Norita Kawanaka}
\affiliation{Department of Physics, Graduate School of Science, Tokyo Metropolitan University, 1-1 Minami-Osawa, Hachioji-shi, Tokyo 192-0397, Japan}
\email{norita@tmu.ac.jp}

\begin{abstract}
In the standard Galactic cosmic-ray (CR) paradigm, protons are accelerated up to $\sim 1$~PeV by Galactic sources. While supernova remnants (SNRs) have been traditionally considered as the primary accelerators, recent observations by LHAASO and HAWC have detected very-high-energy (VHE) gamma rays exceeding 100~TeV from several microquasars, suggesting that these X-ray binaries can accelerate CRs beyond 1~PeV. We investigate the escape process of CRs from microquasars, focusing on the energy-dependent transport mechanisms. High-energy CRs are likely to have long mean free paths and move ballistically on scales smaller than their mean free path, while lower-energy CRs undergo diffusive propagation. This transition results in a spectral break in the CR distribution around the microquasar. We calculate CR energy spectra within a 10--30~pc radius for various diffusion coefficients and timescales. Our model predicts a spectral break and hardening at $E_p \sim 10$--100~TeV when the standard diffusion coefficient for the interstellar space is assumed. However, current VHE gamma-ray observations do not show clear spectral breaks, suggesting that the diffusion coefficient may be significantly reduced near microquasars, possibly due to magnetic field amplification by CR-driven turbulence. 
\end{abstract}


\section{Introduction} 
 \label{sec:intro}

Understanding the origin of Galactic cosmic rays (CRs) remains one of the fundamental challenges in high-energy astrophysics. In the standard paradigm, Galactic CRs are accelerated up to the ``knee'' energy of $\sim 1$~PeV primarily by shock acceleration in supernova remnants (SNRs) \citep{1978ApJ...221L..29B,1983RPPh...46..973D}. Observations with the Fermi satellite and ground-based gamma-ray telescopes have confirmed that SNRs can indeed accelerate CRs up to $\sim 1$~TeV through the detection of pion-decay gamma rays \citep{2008A&A...481..401A,2013Sci...339..807A}. However, both theoretical considerations and recent observations indicate that most SNRs face severe difficulties in accelerating protons up to PeV energies \citep{1983A&A...125..249L,2013MNRAS.431..415B}. The limited magnetic field amplification and finite shock lifetime constrain the maximum energy achievable in typical SNRs \citep{2008ApJ...678..939Z,2023ApJ...958....3D}.

Recent breakthroughs in VHE gamma-ray astronomy have opened new perspectives on PeV CR accelerators. The LHAASO and HAWC observatories have detected gamma rays exceeding 100~TeV from several microquasars---X-ray binaries containing stellar-mass black holes with powerful jets \citep{2018Natur.562...82A,2024Natur.634..557A,2024arXiv241008988L}. These detections strongly suggest that microquasars can accelerate CRs beyond 1~PeV, making them promising candidates for Galactic PeVatrons \citep{2017SSRv..207....5R,2020ApJ...904..188K,2020MNRAS.493.3212C,2021PASJ...73..530S,2025MNRAS.541.2434O,2025arXiv251207781A}.  

The spatial distribution and energy spectrum of CRs around their sources provide crucial diagnostics for understanding acceleration and propagation mechanisms. In the interstellar medium (ISM), CR propagation is generally described by diffusion with an energy-dependent diffusion coefficient, $D(E_p)$, where $E_p$ is the particle energy \citep{1990acr..book.....B}. The coefficient and the mean free path of CRs, $\ell(E_p)$, increases with energy as $D(E_p) \propto \ell(E_p) \propto E_p^{0.3-0.8}$ \citep{2010A&A...516A..67M}. For CRs with energies $\gtrsim 1$~PeV, the mean free path can reach $\sim 100$~pc (see Figure~\ref{fig:mfp}), which is comparable to or even larger than the point spread function (PSF) of current gamma-ray telescopes such as LHAASO ($\sim 0.3^\circ$ at 100~TeV, corresponding to $\sim 10$~pc at $d\sim 2$~kpc distance; \citealt{2024Univ...10..100L}).

A critical consequence of the energy-dependent mean free path is the transition between ballistic and diffusive propagation regimes. On spatial scales smaller than $\ell(E_p)$, CRs move ballistically with negligible scattering, while on larger scales they undergo diffusive transport \citep{2025PhRvD.111l3016K,2025ApJ...982...85L}. This transition should manifest as a spectral break in the observed CR distribution: high-energy CRs in the ballistic regime preserve the source injection spectrum, whereas lower-energy CRs in the diffusive regime exhibit spectra modified by energy-dependent diffusion. The location and shape of this spectral break encode information about the diffusion coefficient and magnetic turbulence near the source.

In this paper, we investigate the spatial distributions and energy spectra of CRs around microquasars, explicitly accounting for the transition between ballistic and diffusive propagation regimes. We calculate CR spectra for various values of the diffusion coefficient reduction factor and discuss implications for current VHE gamma-ray observations. Our analysis provides constraints on CR transport properties near microquasars.

The paper is organized as follows. In Section~\ref{sec:model}, we describe our model assumptions including the source configuration, CR injection spectrum, and diffusion coefficient. Section~\ref{sec:result} presents the calculated CR spectra for different timescales and diffusion coefficients. In Section~\ref{sec:discussion}, we compare our predictions with observations and discuss physical mechanisms for diffusion coefficient reduction. Finally, Section~\ref{sec:conclusion} summarizes our main findings and their implications.

\section{Models} 
\label{sec:model}

\subsection{Cosmic-Ray Injection}

We consider a simplified model in which a microquasar serves as a point source of CRs. We assume spherical symmetry with the microquasar located at $r = 0$. 

We assume a power-law escaping CR spectrum injected per time
\begin{equation}
Q(E_p) dE_p = Q_0 \left(\frac{E_p}{E_{p0}}\right)^{-2} dE_p 
\label{eq:Qp}
\end{equation}
for $t > 0$ and $Q = 0$ for $t < 0$, where $Q_0$ is the normalization constant and $E_{p0}$ is a reference energy. In this study, we are interested only in the shape of the CR spectra, and we do not discuss the values of $Q_0$ and $E_{p0}$. This spectral index of $-2$ is typical for shock acceleration. 

\subsection{Diffusion Coefficient}

The diffusion coefficient for CRs in the ISM is parameterized as
\begin{equation}
D(E_p) = 10^{28}\chi \left(\frac{E_p}{10~{\rm GeV}}\right)^{1/2} ~{\rm cm}^2~{\rm s}^{-1},
\label{eq:Dp}
\end{equation}
where $\chi \leq 1$ is a reduction factor accounting for possible suppression of the diffusion coefficient near the source \citep[e.g.][]{2010ApJ...712L.153F,2011MNRAS.415.3434F,2011MNRAS.410.1577O}.

\begin{figure}[ht!]
\plotone{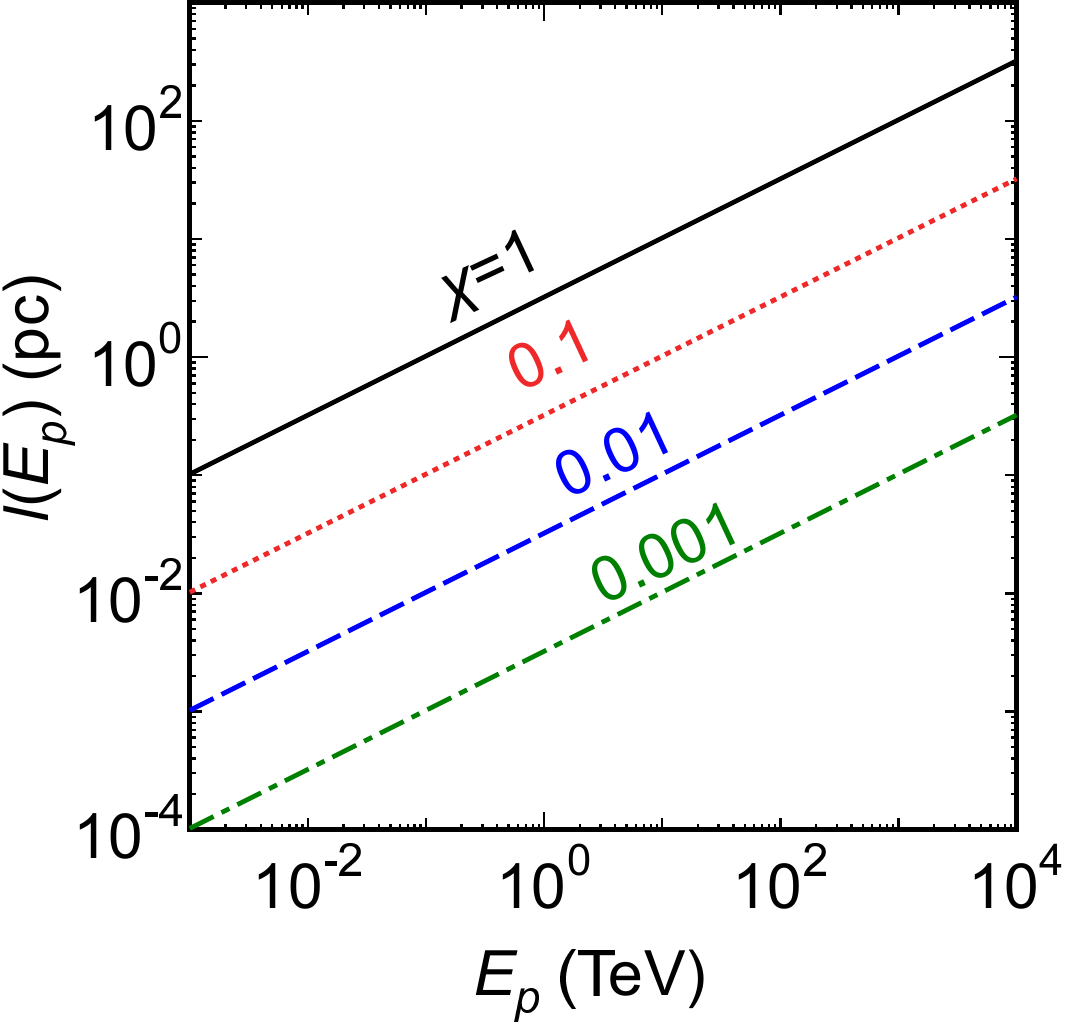}
\caption{The mean free path $\ell(E_p)$ of CRs. The lines represent $\chi=1$ (black solid), 0.1 (red dotted), 0.01 (blue dashed), and 0.001 (green dash-dotted).
\label{fig:mfp}}
\end{figure}

The mean free path of CRs is given by
\begin{equation}
\ell(E_p) = \frac{3D(E_p)}{c} \sim 32\: \chi \left(\frac{E_p}{100~{\rm TeV}}\right)^{1/2}{\rm pc}\:,
\label{eq:mfp}
\end{equation}
where $c$ is the speed of light. Figure~\ref{fig:mfp} shows $\ell(E_p)$ for different $\chi$. The value of $\chi$ critically determines whether a break due to the transition between ballistic and diffusive regimes is observed in the CR spectrum within a given region.

\begin{figure}[ht!]
\plotone{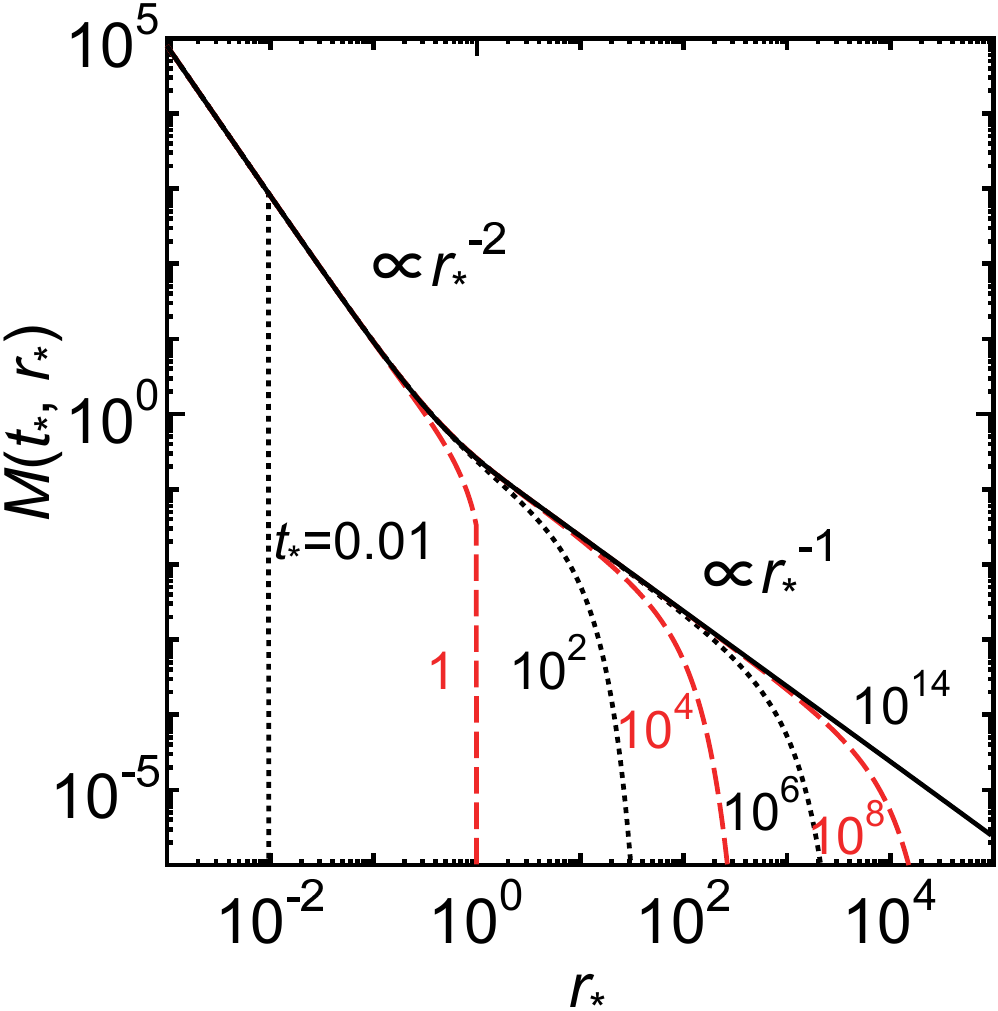}
\caption{The spatial distribution of ultrarelativistic
particles emitted from a particle source at $r_*=0$ with continuous injection starting at $t_* = 0$. The number of particles emitted per unit time is one.  \label{fig:Mtr}}
\end{figure}

\subsection{Analytical Framework}
\label{sec:ana}

To calculate CR distributions around the microquasar, we adopt the analytical solutions for a steady point source derived by \citet{2025PhRvD.111l3016K}. For a particle source with continuous injection starting at $t = 0$, the CR distribution can be expressed in terms of normalized variables $(t_*, r_*) = (ct/\ell, r/\ell)$, where both time and distance are scaled by the energy-dependent mean free path (Figure~\ref{fig:Mtr}).

The transition between ballistic and diffusive regimes occurs when $r \sim \ell(E_p)$, or equivalently $r_* \sim 1$. For $r_* \ll 1$ (ballistic regime), CRs have not experienced significant scattering yet, so they preserve the source injection energy spectrum. The distribution is represented by $\propto r_*^{-2}$ and the particles extend to $r_*\sim t_*$. For $r_* \gg 1$ (diffusive regime), CR transport is well-described by the diffusion equation. The distribution is represented by $\propto r_*^{-1}$ and the particles extend to $r_*\sim \sqrt{t_*}$. The energy spectrum is modified by the energy-dependent diffusion coefficient.

\begin{figure*}[ht!]
\plotone{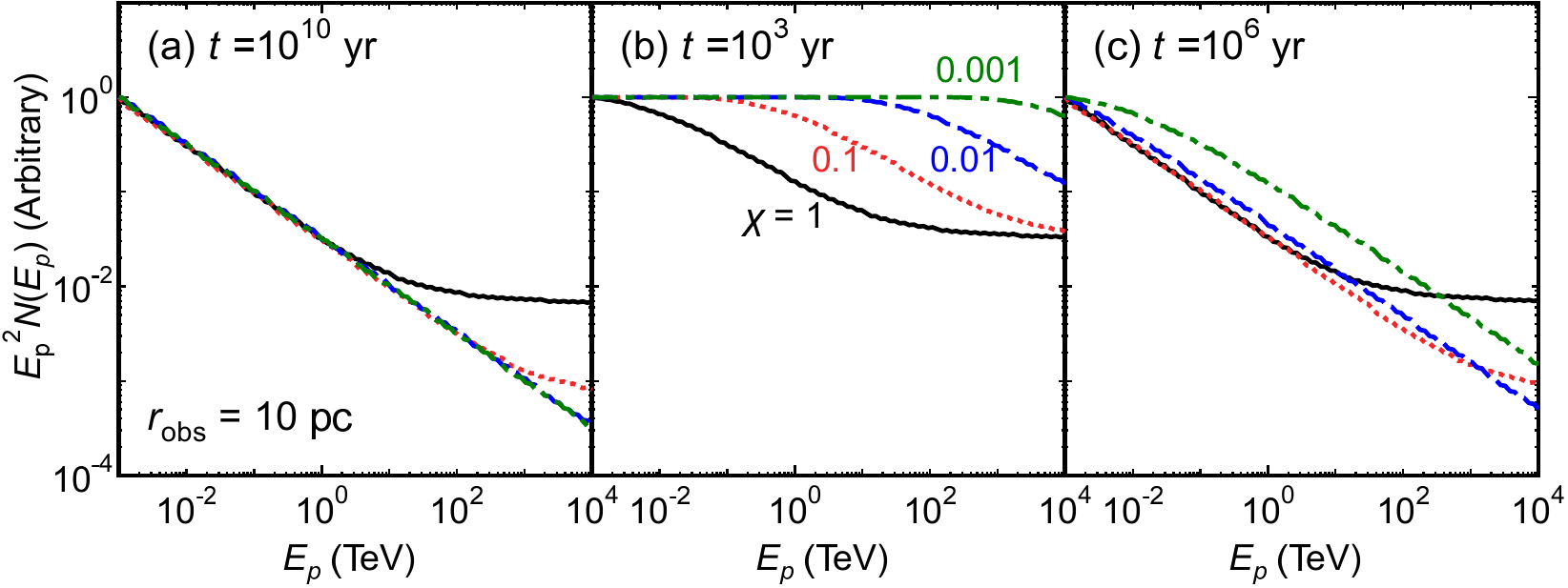}
\plotone{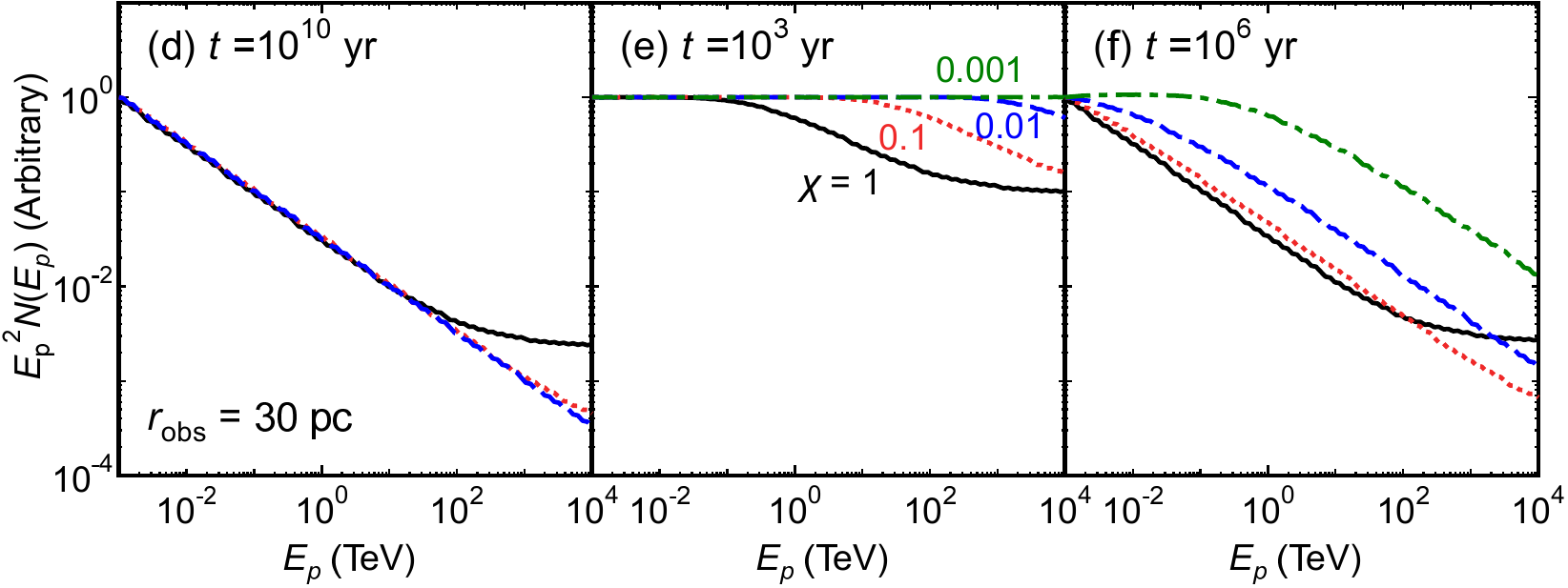}
\caption{The spectra of CRs within $r_{\rm obs}$. The lines represent $\chi=1$ (black solid), 0.1 (red dotted), 0.01 (blue dashed), and 0.001 (green dash-dotted). (a) $r_{\rm obs}=10$~pc, $t=10^{10}$~yr, (b) $r_{\rm obs}=10$~pc, $t=10^{3}$~yr, (c) $r_{\rm obs}=10$~pc, $t=10^{6}$~yr, (d) $r_{\rm obs}=30$~pc, $t=10^{10}$~yr, (e) $r_{\rm obs}=30$~pc, $t=10^{3}$~yr, (f) $r_{\rm obs}=30$~pc, $t=10^{6}$~yr. All lines are normalized at $E_p=1$~GeV.
\label{fig:spec}}
\end{figure*}

\section{Results} 
\label{sec:result}

We calculate the energy spectra of CRs, $N(E_p)$, within an observational radius $r_{\rm obs} = 10$ and 30~pc, which correspond approximately to the PSF of current VHE gamma-ray telescopes ($\sim 0.3^\circ$) at typical microquasar distances of $d\sim 2$ and $\sim 6$~kpc, respectively \citep{2024Univ...10..100L}. Although the typical age of microquasars is $t = 10^6$ yr \citep
[e.g.][]{2021ApJ...910..149O}, we also consider the CR energy spectra at $t = 10^{10}$ yr (almost steady state) and $t = 10^3$ yr (early phase), in addition to $t = 10^6$ yr (typical microquasar age).

\subsection{Almost Steady State ($t = 10^{10}$ yr)}
\label{sec:1010}

Figure~\ref{fig:spec}(a) shows the CR spectra within $r_{\rm obs} = 10$~pc at $t = 10^{10}$~yr, which greatly exceeds typical microquasar ages. At this late time, CRs have diffused far beyond $r_{\rm obs}$ regardless of their energy. 
When $r_{\rm obs} \sim \ell(E_p)$ (i.e., $r_* \sim 1$), the energy is represented by
\begin{equation}
\label{eq:break_a}
 E_p\sim 9.5\: \chi^{-2}\left(\frac{r_{\rm obs}}{10\rm\: pc}\right)^2~\rm TeV
\end{equation}
(see equation~(\ref{eq:mfp})).
Therefore, CRs with $E_p \gtrsim 10$~TeV move ballistically within $r_{\rm obs}$ (ballistic regime), while those with $E_p \lesssim 10$~TeV undergo diffusive transport (diffusive regime).
This regime transition results in a concave spectral break at $E_p \sim 10$~TeV. In the ballistic regime, CR motion is independent of energy and the source spectrum ($\propto E_p^{-2}$; equation~(\ref{eq:Qp})) is preserved. In the diffusive regime, the energy-dependent diffusion coefficient (equation~(\ref{eq:Dp})) modifies the spectrum. The resulting spectral break is a robust prediction when the standard ISM diffusion coefficient is assumed.

For reduced diffusion coefficients with $\chi \lesssim 0.01$, we find $r_{\rm obs} \gtrsim \ell(E_p)$ for $E_p \lesssim 100$~PeV. In this case, all CRs of interest remain in the diffusive regime at the observational scale, and no clear spectral break appears.

Figure~\ref{fig:spec}(d) shows the CR spectra within $r_{\rm obs} = 30$~pc. The break energies are larger than those in Figure~\ref{fig:spec}(a) because of equation~(\ref{eq:break_a}).

\subsection{Early Phase ($t = 10^3$ yr)}
\label{sec:103}

Figure~\ref{fig:spec}(b) shows CR spectra within $r_{\rm obs} = 10$~pc at $t = 10^{3}$~yr. When $\chi$ is small, even the high-energy cosmic rays have not escaped the $r=r_{\rm obs}$ sphere. Thus, the observation captures high-energy CRs without leakage. Consequently, the spectrum does not soften from the injection spectrum. For example, when $\chi = 0.01$, all CRs we consider ($E_p < 10$ PeV) are in the diffusive regime. However, those with $E_p \lesssim 100$~TeV have not yet reached $r = r_{\rm obs}$.
This is because CRs extend to $r_*\sim \sqrt{t_*}$ in the diffusive regime (Section~\ref{sec:ana}) and because $\ell(E_p)\propto \chi E_p^{1/2}$ (equation~(\ref{eq:mfp})). Therefore, CRs with the energies smaller than
\begin{equation}
\label{eq:break_e}
 E_p\sim 100\:\left(\frac{\chi}{0.01}\right)^{-2}\left(\frac{t}{10^3\rm\: yr}\right)^{-2}\left(\frac{r_{\rm obs}}{10\rm\: pc}\right)^4~\rm TeV
\end{equation}
have not reached $r_{\rm obs}$.
Consequently, the spectrum $N(E_p)$ preserves the source spectrum ($\propto E_p^{-2}$) for $E_p \lesssim 100$~TeV, creating a convex spectral break at $E_p \sim 100$~TeV.

For $\chi = 1$, CRs with $E_p \gtrsim 10^{-2}$~TeV have partially diffused beyond $r = r_{\rm obs}$, and those with $E_p \gtrsim 10$~TeV enter the ballistic regime. This produces a convex spectral break at $E_p \sim 10^{-2}$~TeV and a concave break at $E_p \sim 10$~TeV. The resulting spectrum differs qualitatively from that of $t = 10^{10}$~yr.

Figure~\ref{fig:spec}(e) presents the CR spectra within $r_{\rm obs} = 30$~pc. For $\chi = 1$, CRs with $E_p \gtrsim 1$~TeV have partially diffused beyond $r = r_{\rm obs}$. This means that a larger energy is required to reach $r_{\rm obs}$ compared to the case of $r_{\rm obs} = 10$~pc (Figure~\ref{fig:spec}(b)).

\subsection{Typical Microquasar Age ($t = 10^6$ yr)}

In Figure~\ref{fig:spec}(c), we show CR spectra within $r_{\rm obs} = 10$~pc at $t = 10^6$~yr. The spectra represent a mixture of the behaviors seen in Figures~\ref{fig:spec}(a) and~(b). The position of the concave break follows equation~(\ref{eq:break_a}), while the position of the convex break follows equation~(\ref{eq:break_e}).
For the standard diffusion coefficient ($\chi = 1$), a spectral break at $E_p \sim 10$~TeV is clearly visible, transitioning from the diffusive regime at lower energies to the ballistic regime at higher energies.

The spectral shape depends sensitively on $\chi$. For $\chi = 0.1$, the concave break shifts to a higher energy, while for $\chi = 0.01$--$0.001$, the break disappears entirely in the figure. For $\chi = 0.001$, CRs with $E_p \lesssim 10^{-2}$~TeV have not yet reached $r = r_{\rm obs}$. 

Figure~\ref{fig:spec}(f) shows the CR spectra within $r_{\rm obs} = 30$~pc. For $\chi = 1$, the concave break is located at $E_p\sim 100$~TeV. For $\chi = 0.001$, the convex break is found at $E_p \sim 1$~TeV.

This time-dependent evolution reflects the complex interplay between CR diffusion, ballistic propagation, and the finite age of the source.

\section{Discussion} \label{sec:discussion}

\subsection{Comparison with Observations}

The inelastic cross section of pp interactions only weakly depends on particle energy \citep{2006PhRvD..74c4018K}. Since gamma rays created through pp interactions have an energy of $E_\gamma\sim 0.1 E_p$, the spectral break at $E_p$ is observed at $E_\gamma\sim 0.1 E_p$.

Recent LHAASO observations of microquasars such as SS~433 (distance $d=4.6$~kpc, extention at 100~TeV $\theta=0.70^\circ$), V4641~Sgr (6.2~kpc, $0.36^\circ$), GRS~1915+105 (9.4~kpc, $0.33^\circ$), and MAXI J1820+070 (2.96~kpc, $<0.28^\circ$) have detected VHE gamma rays extending to energies beyond $\gtrsim 100$~TeV without showing clear spectral breaks \citep{2024arXiv241008988L}.
These smooth power-law spectra, which do not harden at higher energies, appear to be inconsistent with the predicted concave spectral break at $E_\gamma \sim 10$--$100$~TeV (corresponding to $E_p \sim 100$--$1000$~TeV) expected from our model with the standard or mildly reduced ISM diffusion coefficient assuming $t\sim 10^6$~yr ($\chi \sim 0.1$--1; Figures~\ref{fig:spec}(c) and~(f)). 
We note that the inconsistency is not conclusive given the limited energy range and uncertainty of the current observations.
If the spectral break is actually absent, it would suggest that the diffusion coefficient is significantly reduced near the microquasars, with a value of $\chi \lesssim 0.01$--$0.1$ required to suppress the break in the current observational energy range (Figures~\ref{fig:spec} (c) and (f)). This reduction would place these systems in the diffusive regime, even at PeV energies, resulting in the observed smooth spectra.

\subsection{Physical Mechanisms for Diffusion Coefficient Reduction}

Several physical mechanisms can reduce the diffusion coefficient in the vicinity of microquasars:

\textbf{(1) CR-driven streaming instabilities:} High-intensity CR fluxes can excite Alfv\'en waves through streaming instabilities, amplifying magnetic field turbulence and reducing the CR mean free path \citep{1978MNRAS.182..147B,2009ApJ...707L.179F,2010ApJ...712L.153F}. The growth rate of this instability scales with the CR energy density, which is enhanced near the source.

\textbf{(2) Jet-ISM interaction:} The interaction between the relativistic jet and the ambient ISM could create a complex, highly turbulent environment with amplified magnetic fields \citep{2011MNRAS.414.2838G}. Shocks driven into the ISM may compress and amplify the magnetic field by factors of $\sim 10$--$100$, correspondingly reducing $\chi$ by similar factors.

\subsection{Confinement of Cosmic Rays}

If the diffusion coefficient is significantly reduced around a microquasar, CRs are virtually confined around it.
The escape time for PeV CRs from the reduced-diffusion region (say $r_{\rm diff}\sim 50$~pc) is $t_{\rm esc} \sim r_{\rm diff}^2/D \sim 10^5$--$10^6$~yr for $\chi \sim 0.001$. This timescale is comparable to the typical active lifetime of microquasars. This suggests that a significant fraction of accelerated CRs may be confined near their sources, though they could eventually be released into the ISM.

\section{Conclusions} 
\label{sec:conclusion}

We have investigated the spatial distributions and energy spectra of CRs around microquasars, focusing on the transition between ballistic and diffusive propagation regimes. Our main findings are:

\begin{enumerate}
\item We predict a concave spectral break in the CR distribution at $E_p \sim 10$--100~TeV for the standard ISM diffusion coefficient within a radius of $\sim 10$--30~pc of microquasars with a typical age of $\sim 10^6$~yr. This break arises from the transition between the diffusive regime at lower energies and the ballistic regime at higher energies.

\item The apparent lack of distinct spectral breaks in the current VHE gamma-ray observations of microquasars suggests that the diffusion coefficient is reduced by a factor of $\chi \lesssim 0.01$--$0.1$ near these sources, compared to typical ISM values.

\end{enumerate}

Future observations with wider energy ranges will be useful for detecting the spectral break. Additionally, instruments with improved angular resolution and sensitivity can verify our model predictions. The Cherenkov Telescope Array Observatory (CTAO) will achieve an angular resolution of $\sim 0.05^\circ$ at 1~TeV \citep{2019scta.book.....C}, corresponding to about 2 pc at $d=2$~kpc. This resolution could potentially reveal the spatial extent of VHE gamma-ray emission and detect radial variations in the spectral index predicted by our model.
Our study could also be applied to the interpretation of pulsar gamma-ray halos, which are supposed to be the inverse Compton emission from CR electrons/positrons injected from the pulsars \citep{2017Sci...358..911A,2021PhRvL.126x1103A}.  Since the energies of electrons/positrons that are responsible for these halos are $\gtrsim 100$~TeV, the effects of ballistic propagation can be significant \citep{2021PhRvD.104l3017R}.

\begin{acknowledgments}
This work was supported JSPS KAKENHI grant Nos. 22H00158, 23H04899, 25H00672 (Y.F.), 23K20869, 25K01045 (R.T.) and 22K03686 (N.K.).
\end{acknowledgments}

\bibliography{ms}{}
\bibliographystyle{aasjournalv7}

\end{document}